\documentclass[twocolumn,tighten,times]{aastex631}

\PassOptionsToPackage{dvipsnames,svgnames,x11names}{xcolor}
\usepackage{tabularx}

\usepackage{amsmath}  
\usepackage{amssymb}  
\usepackage[T1]{fontenc}
\usepackage{xspace}
\usepackage{multirow}
\usepackage{xcolor}

\defcitealias{kenworthy_salt3_2021}{K21}

\usepackage{newunicodechar,graphicx}
\DeclareRobustCommand{\okina}{%
  \raisebox{\dimexpr\fontcharht\font`A-\height}{%
    \scalebox{0.8}{`}%
  }%
}
\newunicodechar{ʻ}{\okina}

\newif{\ifchangetext}
\changetextfalse

\InputIfFileExists{\jobname-options}

\ifchangetext
  
  \newcommand{\changenote}[1]{\textcolor{blue}{ \bf #1}}x
\else
  
  \newcommand{\changenote}[1]{}
\fi

\hypersetup{
    colorlinks=True,
    citecolor=black,
    linkcolor=black,
    filecolor=magenta,      
    urlcolor=cyan,
}

\def\arcsec{\ensuremath{^{\prime\prime}}}

\def\xone{\ensuremath{x_{1}}}
\def\C{\ensuremath{\mathcal{C}}}

\newcommand{\STScI}{Space Telescope Science Institute, Baltimore, MD 21218, USA}
\newcommand{\JHU}{Physics and Astronomy Department, Johns Hopkins University, Baltimore, MD 21218, USA}

\newcommand{\ISEF}{ISEF International Fellowship}
\newcommand{\NEF}{NASA Einstein Fellow}
\newcommand{\snz}{2.903}
\newcommand{\mufit}{47.18}
\newcommand{\mufiterrl}{0.28}
\newcommand{\mufiterrh}{0.27}

\usepackage{xspace}

\newcommand{\highzIa}{SN\,2023adsy\xspace}
\usepackage[mathlines]{lineno}
\begin{document}

    \title{Discovery of An Apparent Red, High-Velocity Type Ia Supernova at $\mathbf{z=2.9}$ with \textit{JWST}}

\author[0000-0002-2361-7201
]{J.~D.~R.~Pierel}
\correspondingauthor{J.~D.~R.~Pierel} 
\email{jpierel@stsci.edu}
\altaffiliation{\NEF}
\affiliation{\STScI}

\author[0000-0003-0209-674X]{M.~Engesser}
\affiliation{\STScI}

\author[0000-0003-4263-2228]{D.~A.~Coulter} 
\affiliation{\STScI} 

\author[0000-0002-4781-9078]{C.~DeCoursey}
\affiliation{Steward Observatory, University of Arizona, 933 N. Cherry Avenue, Tucson, AZ 85721 USA}

\author[0000-0003-2445-3891]{M.~R.~Siebert} 
\affiliation{\STScI}

\author[0000-0002-4410-5387]{A.~Rest} 
\affiliation{\STScI}
\affiliation{\JHU}

\author[0000-0003-1344-9475]{E.~Egami}
\affiliation{Steward Observatory, University of Arizona, 933 N. Cherry Avenue, Tucson, AZ 85721 USA}

\author[0000-0003-1060-0723]{W.~Chen} 
\affiliation{Department of Physics,
Oklahoma State University, 145 Physical Sciences Bldg, Stillwater, OK
74078, USA}

\author[0000-0003-2238-1572]{O.~D.~Fox} 
\affiliation{\STScI}

\author[0000-0002-6230-0151]{D.~O.~Jones} 
\affiliation{Institute for Astronomy, University of Hawaiʻi, 640 N. A’ohoku Pl., Hilo, HI 96720, USA}

\author[0000-0002-7593-8584]{B.~A.~Joshi} 
\affiliation{\JHU}

\author[0000-0003-1169-1954]{T.~J.~Moriya}
\affiliation{National Astronomical Observatory of Japan, National Institutes of Natural Sciences, 2-21-1 Osawa, Mitaka, Tokyo 181-8588, Japan}
\affiliation{Graduate Institute for Advanced Studies, SOKENDAI, 2-21-1 Osawa, Mitaka, Tokyo 181-8588, Japan}
\affiliation{School of Physics and Astronomy, Monash University, Clayton, Victoria 3800, Australia}

\author[0000-0002-0632-8897]{Y.~Zenati}
\altaffiliation{\ISEF}
\affiliation{\JHU}
\affiliation{\STScI}

\author[0000-0002-8651-9879] {A.~J.~Bunker}
\affiliation{Department of Physics, University of Oxford, Denys Wilkinson Building, Keble Road, Oxford OX1 3RH, UK}

\author[0000-0002-1617-8917] {P.~A.~Cargile}
\affiliation{Center for Astrophysics $|$ Harvard \& Smithsonian, 60 Garden St., Cambridge MA 02138 USA}

\author[0000-0002-2678-2560] {M.~Curti}
\affiliation{European Southern Observatory, Karl-Schwarzschild-Strasse 2, 85748 Garching, Germany}

\author[0000-0002-2929-3121] {D.~J.~Eisenstein}
\affiliation{Center for Astrophysics $|$ Harvard \& Smithsonian, 60 Garden St., Cambridge MA 02138 USA}

\author[0000-0003-3703-5154]{S.~Gezari}
\affiliation{\STScI}

\author[0000-0001-6395-6702]{S.~Gomez}
\affiliation{\STScI}

\author[0000-0002-5063-0751]{M.~Guolo} 
\affiliation{\JHU}

\author[0000-0002-9280-7594] {B.~D.~Johnson}
\affiliation{Center for Astrophysics $|$ Harvard \& Smithsonian, 60 Garden St., Cambridge MA 02138 USA}

\author[0000-0003-2495-8670]{M.~Karmen} 
\affiliation{\JHU}

\author[0000-0002-4985-3819] {R.~Maiolino}
\affiliation{Kavli Institute for Cosmology, University of Cambridge, Madingley Road, Cambridge CB3 0HA, UK}
\affiliation{Cavendish Laboratory, University of Cambridge, 19 JJ Thomson Avenue, Cambridge CB3 0HE, UK}
\affiliation{Department of Physics and Astronomy, University College London, Gower Street, London WC1E 6BT, UK}

\author[0000-0001-9171-5236]{Robert~M.~Quimby}
\affiliation{Department of Astronomy/Mount Laguna Observatory, SDSU, 5500 Campanile Drive, San Diego, CA 92812-1221, USA}
\affiliation{Kavli Institute for the Physics and Mathematics of the Universe (WPI), The University of Tokyo Institutes for Advanced Study, The University of Tokyo, Kashiwa, Chiba 277-8583, Japan}

\author[0000-0002-4271-0364] {B.~Robertson}
\affiliation{Department of Astronomy \& Astrophysics, University of California, Santa Cruz, 1156 High Street, Santa Cruz CA 96054, USA}

\author[0000-0002-9301-5302]{M.~Shahbandeh} 
\affiliation{\STScI}

\author[0000-0002-7756-4440]{L.~G.~Strolger} 
\affiliation{\STScI}

\author[0000-0002-4622-6617]{F.~Sun}
\affiliation{Center for Astrophysics $|$ Harvard \& Smithsonian, 60 Garden St., Cambridge MA 02138 USA}

\author[0000-0001-5233-6989]{Q.~Wang} 
\affiliation{\JHU}

\author[0000-0002-4043-9400]{T.~Wevers}
\affiliation{\STScI}


\begin{abstract}
We present the \textit{JWST} discovery of \highzIa, a transient object located in a host galaxy JADES-GS$+53.13485$$-$$27.82088$ with a host spectroscopic redshift of $\snz\pm0.007$. The transient was identified in deep \textit{James Webb Space Telescope} (\textit{JWST})/NIRCam imaging from the \textit{JWST} Advanced Deep Extragalactic Survey (JADES) program. Photometric and spectroscopic followup with NIRCam and NIRSpec, respectively, confirm the redshift and yield UV-NIR light-curve, NIR color, and spectroscopic information all consistent with a Type Ia classification. Despite its classification as a likely SN\,Ia, \highzIa is both fairly red ($E(B-V)\sim0.9$) despite a host galaxy with low-extinction and has a high Ca\,II velocity ($19,000\pm2,000$ km/s) compared to the general population of SNe\,Ia. While these characteristics are consistent with some Ca-rich SNe\,Ia, particularly SN\,2016hnk, \highzIa is intrinsically brighter than the low-$z$ Ca-rich population. Although such an object is too red for any low-$z$~cosmological sample, we apply a fiducial standardization approach to \highzIa~and find that the \highzIa luminosity distance measurement is in excellent agreement ($\lesssim1\sigma$) with $\Lambda$CDM. Therefore unlike low-$z$ Ca-rich SNe\,Ia, \highzIa is standardizable and gives no indication that SN\,Ia standardized luminosities change significantly with redshift. A larger sample of distant SNe\,Ia is required to determine if SN\,Ia population characteristics at high-$z$ truly diverge from their low-$z$ counterparts, and to confirm that standardized luminosities nevertheless remain constant with redshift.

\pagebreak
\end{abstract}

\section{Introduction}
\label{sec:intro}
Type Ia supernovae (SNe\,Ia) have now been used for decades as precise luminosity distance measures, enabling the discovery of dark energy and our best local measurement of the Hubble constant \citep[$H_0$;][]{riess_observational_1998,perlmutter_measurements_1999,riess_comprehensive_2022}. SNe\,Ia can be found over a wide redshift range, making them an ideal tool for measuring changes in dark energy over time. However, doing so requires 1) a large sample of well-observed SNe\,Ia and 2) that the standardization properties of SNe\,Ia do not change with redshift. This second point is particularly important, as many redshift-evolving global properties could plausibly impact SN\,Ia luminosities and mimic the signal of evolving dark energy \citep[e.g., metallicity;][]{moreno-raya_dependence_2016}. This effect could bias dark energy measurements below the level of our current measurement precision \citep{riess_first_2006,scolnic_complete_2018,brout_pantheon_2022}.

The exact nature of dark energy is one of the fundamental questions for cosmology, and next-generation SN\,Ia dark energy measurements will rely upon SN\,Ia luminosities remaining constant with redshift to remain unbiased. Evolving luminosity distances could indicate dark energy and/or SN\,Ia intrinsic luminosity are changing with redshift, making it difficult to distinguish between the two effects. In the dark-matter dominated universe beyond $z\sim2$, dark energy variation is expected to be very small, and so evolution in luminosity distances would strongly indicate intrinsic SN\,Ia luminosity evolution, giving high-$z$ SNe\,Ia unique leverage on SN\,Ia systematics \citep{riess_first_2006}. 

Distance measurements for SNe\,Ia have been made to $z=2.22$ with the \textit{Hubble Space Telescope} \citep[\textit{HST;}][]{rodney_type_2014}, but considering only spectroscopically confirmed SNe\,Ia with spectroscopic redshifts that have not been gravitationally lensed \citep[which adds many systematics, see][]{pierel_jwst_2024} that sample is limited to $z\sim1.6$ \citep{riess_type_2018}. There are only five SNe\,Ia with luminosity distance measurements in the range $1.6<z<2.22$, with two gravitationally lensed \citep{jones_discovery_2013,rubin_discovery_2018} and three photometrically classified \citep{rodney_type_2014}. Two additional spectroscopically confirmed, gravitationally lensed SNe\,Ia have been found at $z=1.78$ \citep{polletta_spectroscopy_2023,chen_jwst_2024,frye_jwst_2024,pascale_sn_2024,pierel_jwst_2024} and $z=1.95$ \citep{pierel_lensed_2024} but they lack luminosity distance measurements. Detecting SNe\,Ia at $z>2$ requires deep (m$_{AB}\gtrsim26$) imaging observations in red ($\gtrsim1.5\mu$m) filters, while spectroscopic confirmation of SNe\,Ia at $z>2$ requires similar depths at wavelengths beyond $\sim2\mu$m to identify the characteristic SiII feature \citep[e.g.,][]{filippenko_optical_1997}. These combined requirements have been beyond the reach of modern observatories until the launch of the \textit{James Webb Space Telescope} (\textit{JWST}). \textit{JWST} has expanded our view of the universe to remarkable distances, and despite a relatively small field of view (FoV) it has been highly efficient at detecting rare SNe at high-$z$ due to its sensitivity and wavelength coverage \citep{engesser_detection_2022,engesser_discovery_2022, decoursey_discovery_2023,decoursey_discovery_2023-1,decoursey_discovery_2023-2,pierel_jwst_2024, pierel_lensed_2024}.

A candidate for the most distant SN\,Ia yet discovered has been found in \textit{JWST} imaging conducted as part of the \textit{JWST} Advanced Deep Extragalactic Survey (JADES) program \citep{eisenstein_jades_2023}. JADES observed $\sim25\arcmin^2$ of sky to extreme depths ($m_{AB}>30$ in $9$ filters) in November 2022 and again in November 2023, giving a sufficiently long baseline to search for transient objects with sensitivity for SNe\,Ia to $z>4$. Of the dozens of detected transient objects, one (subsequently named \highzIa and found in the galaxy JADES-GS$+53.13485$$-$$27.82088$ at R.A.$=3$h$32$m$32.3647$s decl.$=-27$d$49$m$15.238$s), was identified by first-epoch colors, redshift, and luminosity as a possible SN\,Ia candidate at $z\sim2.8$.  A \textit{JWST} Director's Discretionary Time (DDT) program was approved to follow-up the most interesting transients in the field \citep{egami_jwst_2023}, providing two additional imaging epochs and a spectrum for $\sim10$ SNe including \highzIa, which received a refined spectroscopic redshift of $z=\snz\pm0.007$.

While the overall JADES$+$DDT observations and SN population analysis are presented in a companion paper \citep[][, hereafter D24]{decoursey_jades_2024}, here we describe the classification and analysis of \highzIa in detail. We begin by a summary of the observations in Section \ref{sec:obs}, followed by a description of the classification for \highzIa using both the spectrum and light curve in Section \ref{sec:class}. Light curve fitting and the subsequent standardized distance measurement are completed in Section \ref{sec:distance}, and we conclude in Section \ref{sec:conclusion} with prospects for the future of high-$z$ SN\,Ia cosmology, and the implications of the new frontier enabled by \textit{JWST}. In this analysis, we assume a standard flat $\Lambda$CDM cosmology with $H_0=70$km s$^{-1}$ Mpc$^{-1}$, $\Omega_m=0.315$.

\begin{figure*}[th!]
    \centering
    \includegraphics[width=\textwidth,trim={0cm 6.25cm 4.3cm 0cm},clip]{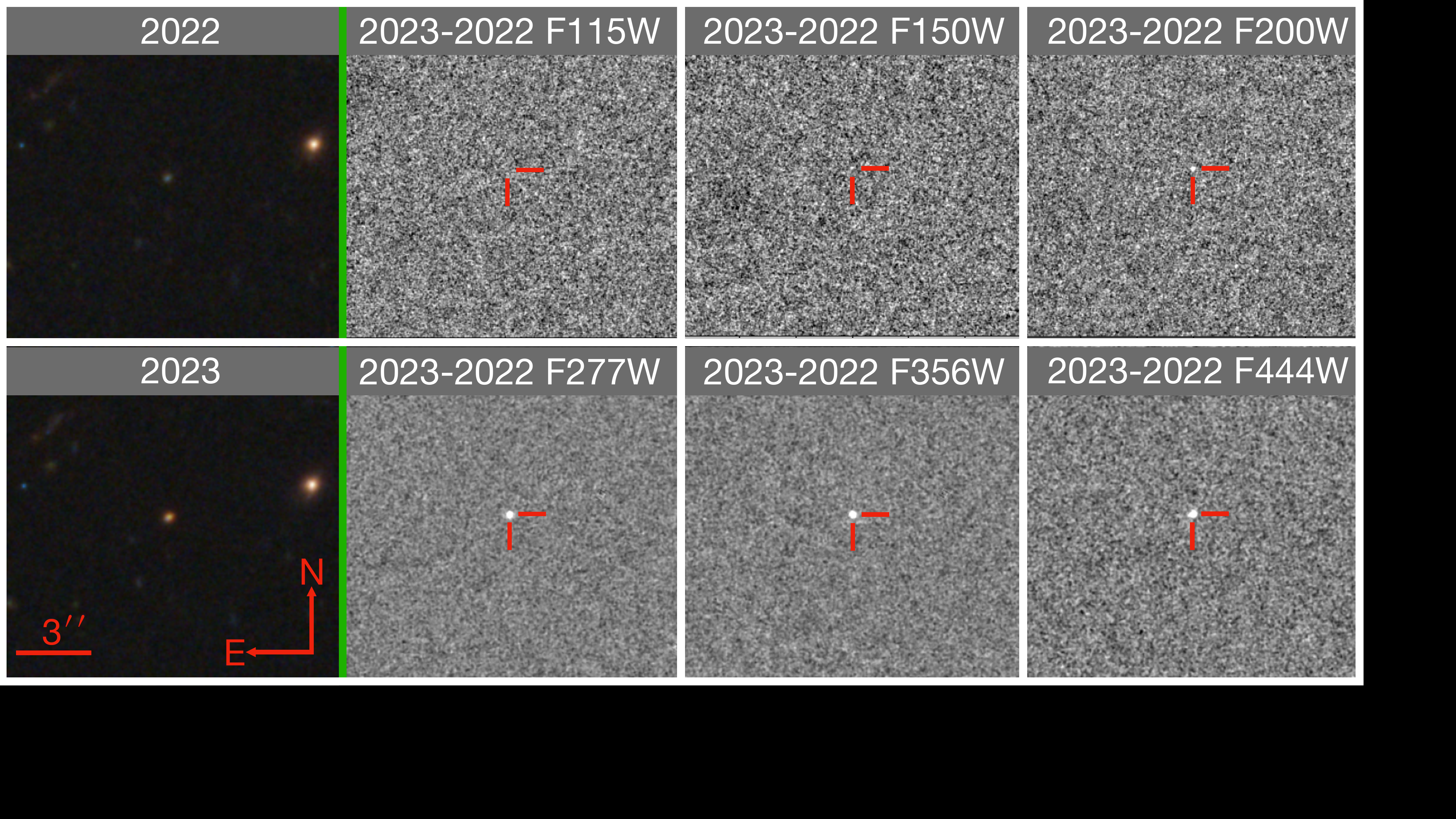}
    \caption{(Left column) Full color images using F115W+F150W (Blue) F200W+F277W (Green) and F356W+F444W (Red), with the $2022$ JADES epoch on top and $2023$ (including \highzIa) on the bottom. (Column $2$-$4$) Difference images created from the two JADES epochs ($2023-2022$), with the \highzIa position marked with a red indicator. All images are drizzled to $0.03^{\prime\prime}/$pix and have the same spatial extent. }
    \label{fig:im_cutouts}
\end{figure*}

\section{Summary of Observations}
\label{sec:obs}
The description of JADES, its observing strategy and the resulting $2022$-$2023$ data products, the method for detecting SNe, and the subsequent DDT program observations are described in detail by D24. Briefly, the initial JADES observations (PID $1180$) were taken over the observing window $2022$ September $29$-October $5$, and the second epoch took place between $2023$ September $29$-October $3$ with an overlap of $25\arcmin^2$ and a $5\sigma$ depth of m$_{AB}\sim30$ in the NIRCam F090W, F115W, F150W, F200W, F277W, F335M, F356W, F410M, F444W filters. There are also additional visits on $2023$ November $15$ and $2024$ January $1$ due to failed observations. A \textit{JWST} DDT program (PID $6541$) was approved to follow the most interesting transients identified with two additional NIRCam visits on $2023$ November $28$ and $2024$ January $1$, with the latter visit including seven hours of integration in the NIRSpec \citep{jakobsen_near-infrared_2022} multi-object spectroscopy (MOS) mode using the micro-shutter assembly \citep[MSA;][]{ferruit_near-infrared_2022} and Prism (R$\sim100$). The MSA provided SN spectra for $\sim10$ transients, most described in companion papers \citep[e.g., D.~Coulter et al. in preparation,][]{siebert_discovery_2024} as well as a variety of galaxy spectra. Below we describe the data reduction and analysis for \highzIa.

\subsection{Measuring Photometry}
\label{sub:obs_phot}
As described in detail by D24, we adopt the point-spread function (PSF) fitting method developed in \citet{pierel_jwst_2024} for measuring photometry on Level 3 (drizzled, I2D) \textit{JWST} images. Unlike their scenario though, we have a template image for all epochs of \highzIa from the $2022$ JADES observations. We therefore first align the Level 2 (CAL) NIRCam images containing \highzIa to the Level 3 template images (I2Ds\footnote{\url{https://archive.stsci.edu/hlsp/jades}}, in each filter) using the \textit{JWST}/\textit{HST} Alignment Tool \citep[{\tt JHAT};][]{rest_arminrestjhat_2023}\footnote{\url{https://jhat.readthedocs.io}}) software and then produce aligned Level 3 images with the \textit{JWST} pipeline \citep{bushouse_jwst_2022}. JHAT improves the relative alignment from  $\sim1$pixel to $\sim0.1$pixel between the epochs. We obtain difference images in all filters using the High Order Transform of PSF and Template Subtraction \citep[{\tt HOTPANTS};][]{becker_hotpants_2015}\footnote{\url{https://github.com/acbecker/hotpants}}) code \citep[with modifications implemented in the \texttt{photpipe} code;][]{rest_testing_2005}, with all short- and long-wavelength (SW and LW, respectively) first-epoch filters shown in Figure \ref{fig:im_cutouts}. We then implement the \texttt{space\_phot}\footnote{\url{space-phot.readthedocs.io}} Level 3 PSF fitting routine from \citet{pierel_jwst_2024} using $5\times5$ pixel cutouts and PSF models from {\tt webbpsf}\footnote{\url{https://webbpsf.readthedocs.io}}, which are temporally and spatially dependent and include a correction to the infinite aperture flux. These total fluxes, which are in units of MJy/sr, are converted to AB magnitudes using the native pixel scale of each image ($0.03\arcsec/$pix for SW, $0.06\arcsec/$pix for LW). Measured photometry is given in Table \ref{tab:phot}.

\begin{table}
    \centering
    \caption{\label{tab:phot} Observations for \highzIa discussed in Section \ref{sec:obs}.}
    
    \begin{tabular*}{\linewidth}{@{\extracolsep{\stretch{1}}}*{5}{c}}
\toprule
PID&MJD&Instrument&\multicolumn{1}{c}{Filter/Disperser}&\multicolumn{1}{c}{m$_{AB}$}\\
\hline
$1180$&$60220$&NIRCam&F090W&$>30.2$\\
$1180$&$60220$&NIRCam&F115W&$>30.6$\\
$1180$&$60220$&NIRCam&F150W&$30.39\pm0.18$\\
$1180$&$60221$&NIRCam&F200W&$28.98\pm0.09$\\
$1180$&$60220$&NIRCam&F277W&$28.26\pm0.05$\\
$1180$&$60220$&NIRCam&F335M&$28.00\pm0.07$\\
$1180$&$60220$&NIRCam&F356W&$28.10\pm0.06$\\
$1180$&$60220$&NIRCam&F410M&$28.07\pm0.08$\\
$1180$&$60220$&NIRCam&F444W&$28.06\pm0.07$\\
\hline
$1180$&$60264$&NIRCam&F090W&$>29.9$\\
$1180$&$60264$&NIRCam&F115W&$>30.3$\\
$1180$&$60264$&NIRCam&F150W&$>30.1$\\
$1180$&$60264$&NIRCam&F200W&$29.00\pm0.12$\\
$1180$&$60264$&NIRCam&F277W&$28.41\pm0.08$\\
$1180$&$60264$&NIRCam&F335M&$28.13\pm0.09$\\
$1180$&$60264$&NIRCam&F356W&$28.45\pm0.09$\\
$1180$&$60264$&NIRCam&F410M&$28.57\pm0.17$\\
$1180$&$60264$&NIRCam&F444W&$28.21\pm0.11$\\
\hline
$6541$&$60276$&NIRCam&F115W&$>28.9$\\
$6541$&$60276$&NIRCam&F150W&$>29.6$\\
$6541$&$60276$&NIRCam&F200W&$28.86\pm0.18$\\
$6541$&$60276$&NIRCam&F277W&$28.53\pm0.15$\\
$6541$&$60276$&NIRCam&F356W&$28.49\pm0.16$\\
$6541$&$60276$&NIRCam&F444W&$28.67\pm0.29$\\
\hline
$1180$&$60311$&NIRCam&F090W&$>29.4$\\
$1180$&$60311$&NIRCam&F115W&$>29.9$\\
$6541$&$60310$&NIRCam&F150W&$>29.9$\\
$6541$&$60310$&NIRCam&F200W&$29.33\pm0.19$\\
$6541$&$60310$&NIRCam&F277W&$28.46\pm0.15$\\
$1180$&$60311$&NIRCam&F335M&$28.26\pm0.10$\\
$6541$&$60310$&NIRCam&F356W&$28.47\pm0.17$\\
$1180$&$60311$&NIRCam&F410M&$>28.4$\\
$6541$&$60310$&NIRCam&F444W&$>28.9$\\
\hline
$6541$&$60310$&NIRSpec&Prism&--\\
\hline
\hline
    \end{tabular*}
\begin{flushleft}
\tablecomments{Columns are: \textit{JWST} Program ID, Modified Julian date, \textit{JWST} instrument, filter or grating, and photometry plus final uncertainty for \highzIa. Upper limits are $2\sigma$.}

\end{flushleft}
\end{table}

\subsection{NIRSpec Reduction}
\label{sub:obs_spec}

\begin{figure*}
    \centering
    \includegraphics[width=\textwidth,trim={0cm 0cm 0cm 0cm},clip]{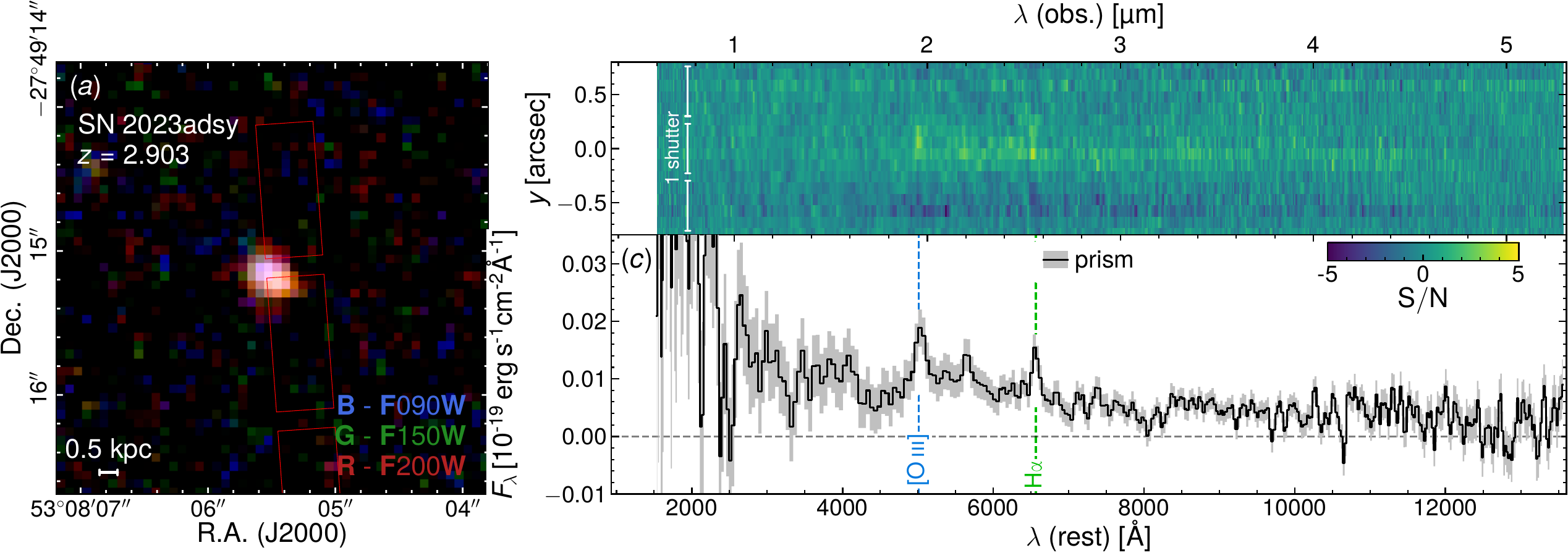}
    \caption{(a) The slitlet positions over \highzIa for one of the dithered observations. The reported slitlet position (shown) is slightly offset from its true position, which we confirm contains the SN. (b/c) The 2D and 1D-extracted NIRSpec spectrum for \highzIa. The primary host emission lines ([O\,III] and H$\alpha$) used for the spectroscopic redshift measurement are shown with dotted lines.  }
    \label{fig:spec_2d_cuts}
\end{figure*}

We began processing the spectroscopic data with Stage 2 products from the Mikulski Archive for Space Telescopes (MAST). Additional processing used the \textit{JWST} pipeline \citep[v$1.12.5$;][]{bushouse_jwst_2022} with context file jwst\_1183.pmap to produce two-dimensional (2D) spectral data (Figure \ref{fig:spec_2d_cuts}). The pipeline applied a slit-loss throughput correction for \highzIa based on the planned position of a point-source within the MSA shutters (Figure \ref{fig:spec_2d_cuts}). The spectra of the SN and its host galaxy was extracted using the
optimal extraction algorithm from \citet{horne_optimal_1986} implemented as scripts available as part of the MOS Optimal Spectral Extraction (MOSE) notebook\footnote{\url{https://spacetelescope.github.io/jdat_notebooks/notebooks/ifu_optimal/ifu_optimal.html}}. We used {\tt webbpsf} to generate the PSF for the NIRSpec observation. As there is no obvious extended emission from the host galaxy, we used a Gaussian kernel to model the flux distribution in the 2D spectrum. The raw Host$+$SN spectrum is shown in the bottom of Figure \ref{fig:spec_2d_cuts}, used for the spectroscopic redshift measurement, and the final SN spectrum used for classification is shown and analyzed in Figure \ref{fig:sn_spec}.

\section{Classification as Type Ia}
\label{sec:class}
\subsection{Spectroscopic Classification}
\label{sub:class_spec}
The first step for our classification is to obtain a spectroscopic redshift by identifying host galaxy emission lines. The two most prominent features are best-matched by [O III] and H$\alpha$, which have rest-frame wavelengths of $\sim5008\AA$ and $\sim6565\AA$ and provide a robust spectroscopic redshift of $z=\snz\pm0.007$ for \highzIa (Figure \ref{fig:spec_2d_cuts}). We use this value for all analysis going forward.

\begin{figure}
    \centering
    \includegraphics[width=.5\textwidth,trim={1.75cm 1.25cm 1.25cm 1.5cm},clip]{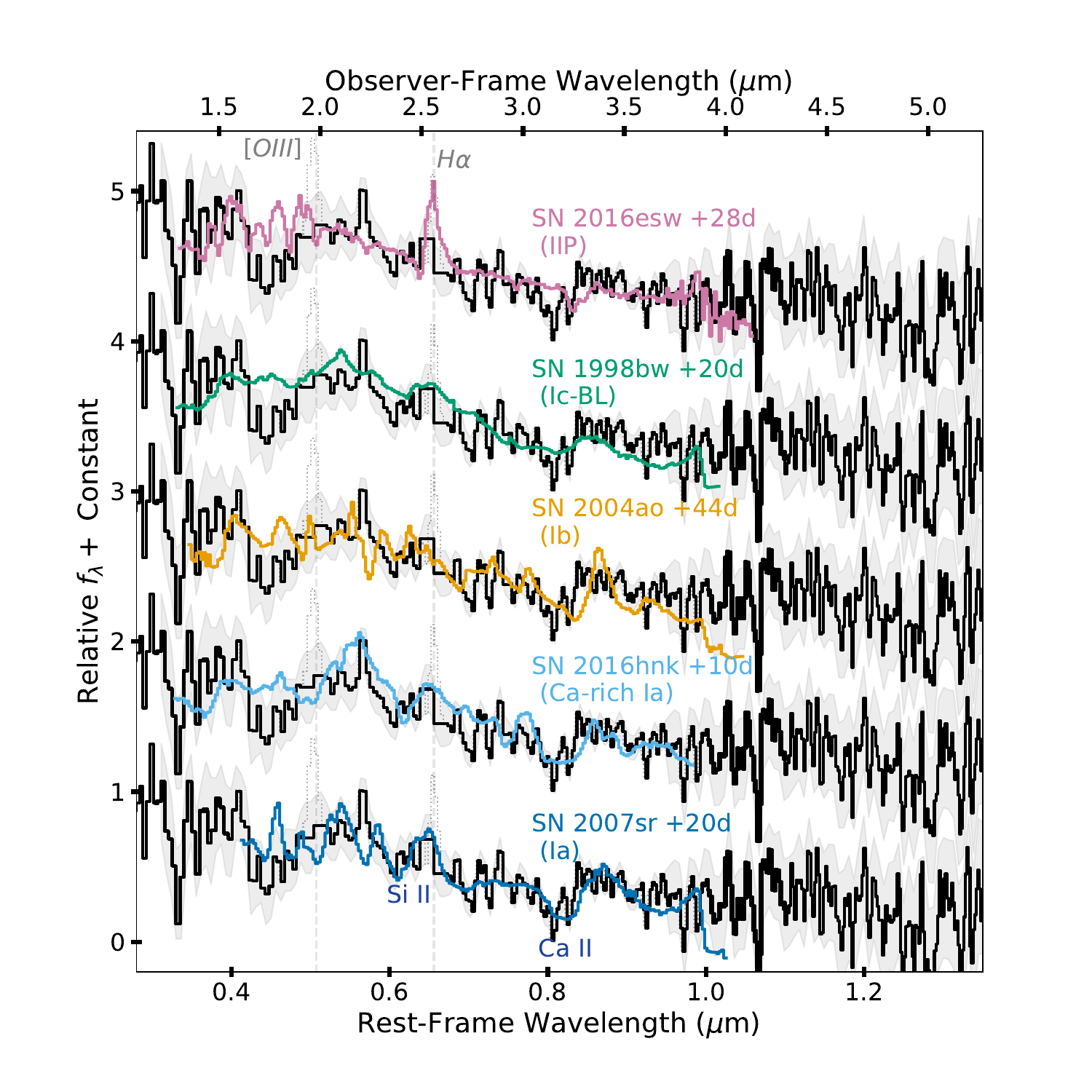}
        \caption{The observed NIRSpec spectrum (with uncertainty) of \highzIa is shown as a black solid line, with the primary features used for the preferred SN\,Ia classification labeled (bottom). The observed host galaxy emission lines (faint dotted lines) are marked and have been removed from the spectrum, with a redshift of $z=\snz$ applied. The best-match template from \texttt{NGSF} is a SN\,Ia (blue, bottom), and a Ca-rich SN\,Ia subclass is also shown for comparison. The best core-collapse matches are also shown, including Ib (red, third from bottom), Ic (pink, second from top), and IIP (light blue, top). While SN\,Ia is favored based on the spectrum, we also use the photometry and host galaxy information to make the final classification.}
    \label{fig:sn_spec}
\end{figure}

\begin{figure}
    \centering
    \includegraphics[width=.37\textwidth,trim={0cm .25cm 0cm .25cm},clip]{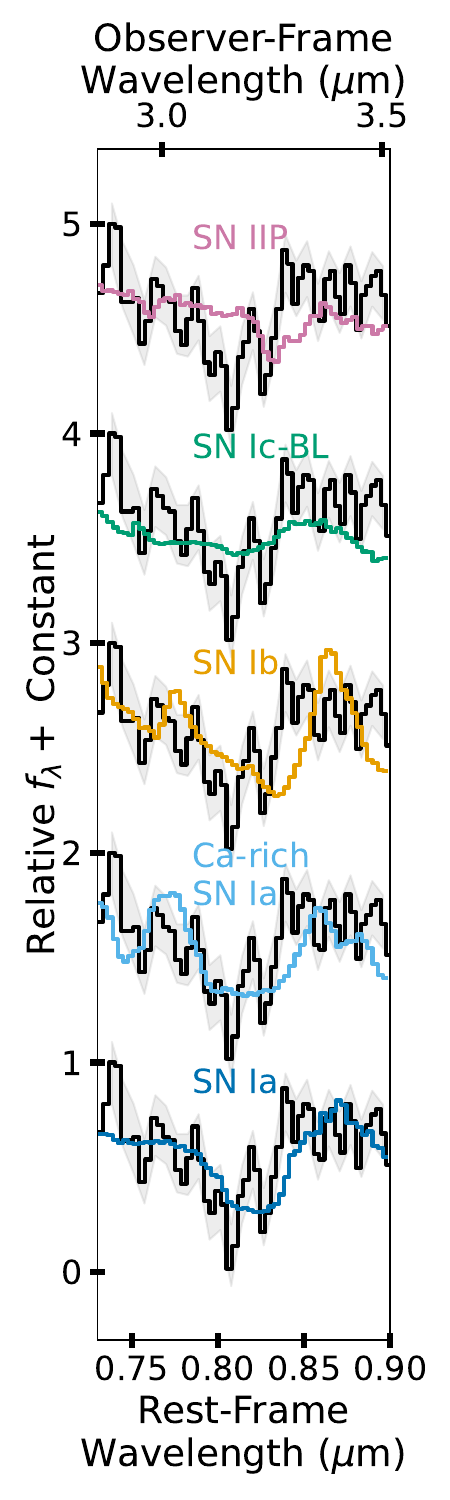}
        \caption{The same as Figure \ref{fig:sn_spec}, but zoomed in on the Ca\,II feature, where the SN\,Ia and Ca-rich subclass are the best fit. }
    \label{fig:sn_spec_ca}
\end{figure}

Next, we remove the host galaxy emission lines from the spectrum and use the Next Generation SuperFit \citep[\texttt{NGSF}][]{goldwasser_next_2022}\footnote{\href{https://github.com/oyaron/NGSF}{https://github.com/oyaron/NGSF}} package to classify \highzIa. We note that by removing the H$\alpha$ emission line we could plausibly be removing SN flux if \highzIa were of Type II, but the width of the line is precisely at the resolution of the Prism (i.e., $\sim3,000$km/s) indicating a narrow emission line consistent with low-velocity host emission. We are therefore confident we are removing H$\alpha$ exclusively from the host galaxy, as a contribution from the SN would result in a line width much broader than what is observed (see the comparison to the SN\,IIP 2016esw Figure \ref{fig:sn_spec}). A narrow emission line could be seen from a SN\,IIn, but the best-fit SN\,IIn \texttt{NGSF} match to the pre-clipped spectrum (Figure \ref{fig:spec_2d_cuts}) results in a $\chi^2/\nu=2.16$, which is still worse than the SN\,Ia match (see below). Additionally, SN\,IIn relative rates are much lower than the SN sub-types we are using for comparison \citep[only $\sim5\%$ of SNe;][]{li_nearby_2011}, making such a discovery very unlikely. Of the top ten reference SN spectra matched to the \highzIa spectrum, $6$ are of Type Ia and the remainder are core-collapse (CC) sub-types, with the best match being Type Ia (Figure \ref{fig:sn_spec} and Table \ref{tab:phase}). The SN\,Ia spectral template match provides a $\chi^2$ per degree of freedom ($\nu$) of $1.72$, while the next best fit is a SN\,Ic with $1.92$ (all values are given in Table \ref{tab:phase}). The primary features being matched are the $6150\AA$ Si\,II and $8300\AA$ Ca\,II absorptions, which are present in the template SN\,Ia spectrum and \highzIa but either not present (Si\,II) or not well-matched (Ca\,II) in the CC spectral templates. The Ca\,II feature is the strongest in the spectrum, with a measured velocity of $\sim19,000\pm2,000$km/s (Figure \ref{fig:sn_spec_ca}). This is relatively high compared to average low-$z$ SNe\,Ia, about $1$-$2\sigma$ above of the observed distribution \citep{siebert_investigating_2019, siebert_asymmetric_2023}, but consistent with a Ca-rich SN (see Section \ref{sub:class_ca-rich}). Given the phase (relative to peak B-band brightness) of the best-fit spectral template for each SN sub-type, the inferred observer-frame times of peak B-band brightness are given in Table \ref{tab:phase} alongside the reduced-$\chi^2$ values and compared to the results from light curve fitting in Section \ref{sub:class_phot}. 

\subsection{Photometric Classification}
\label{sub:class_phot}
While the spectroscopic template matching from the previous section suggests that \highzIa is indeed a SN\,Ia with a best-fit $\chi^2/\nu=1.72$, there is still a possibility that \highzIa is a CC\,SN given the spectrum alone as all best-fit CC\,SN spectral matches have reasonable $1.92<\chi^2/\nu<2.5$. We fit the measured photometry with the SALT3-NIR SN\,Ia light curve model \citep[][and see Section \ref{sub:distance_lc}]{pierel_salt3nir_2022} and all existing CC\,SN light curve evolution models with rest-frame optical to near-IR (to observer-frame $\sim4\mu$m) wavelength coverage \citep{pierel_extending_2018}. We include Galactic dust based on the maps of \citet{schlafly_measuring_2011} and the reddening law from \citet{fitzpatrick_correcting_1999}, which corresponds to $E(B-V)=0.01$mag with $R_V=3.1$. We also allow for a large amount (up to $E(B-V)=1.5$ with $1<R_V<5$) of rest-frame, host-galaxy dust in the CC\,SN light curve fits and a SALT3-NIR color parameter range of $-1.5<c<1.5$ given the very red observed colors. 

\begin{figure}
    \centering
    \includegraphics[width=.5\textwidth,trim={0cm .5cm 1cm 1.8cm},clip]{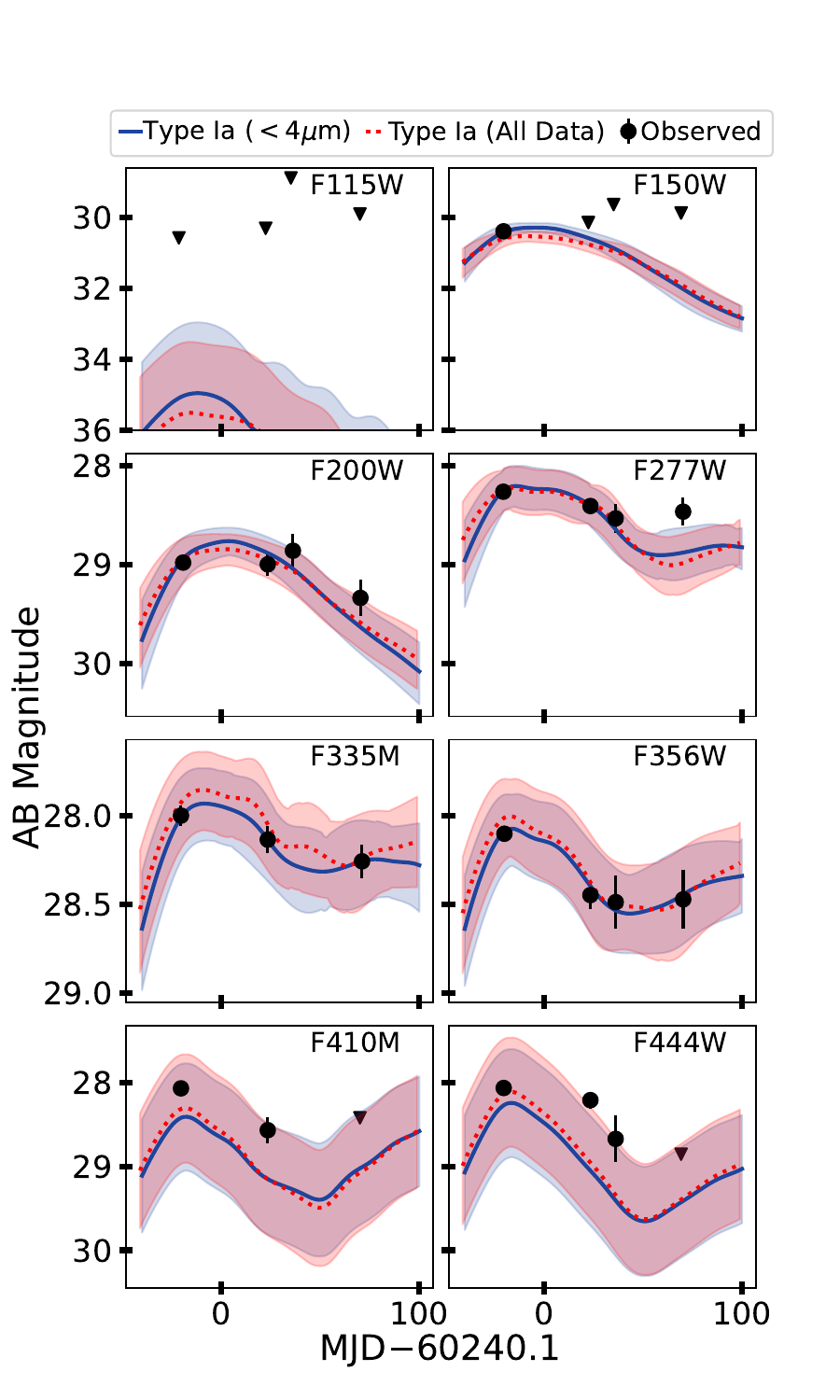}
    \caption{The photometry measured in Section \ref{sub:obs_phot} is shown as black circles with error, with ($5\sigma$) upper-limits denoted by triangles. The fit to the full light curve is shown in red (dashed, with error) and the fit to data $<4\mu$m (i.e., without F410M and F444W) is shown in blue (solid, with error). While the resulting model prediction for the $>4\mu$m data remains roughly the same, the model at $<4\mu$m becomes biased as it attempts to vary the parameters extremely to better fit the reddest data. While the fits are of similar quality, we use the fit to the $<4\mu$m data for our distance modulus measurement as all filters are well-fit without resulting to extreme parameter values (Section \ref{sub:distance_lc}). }
    \label{fig:lc_fit}
\end{figure}

\begin{figure}
    \centering
    \includegraphics[width=.5\textwidth,trim={0cm .5cm 1cm 1.1cm},clip]{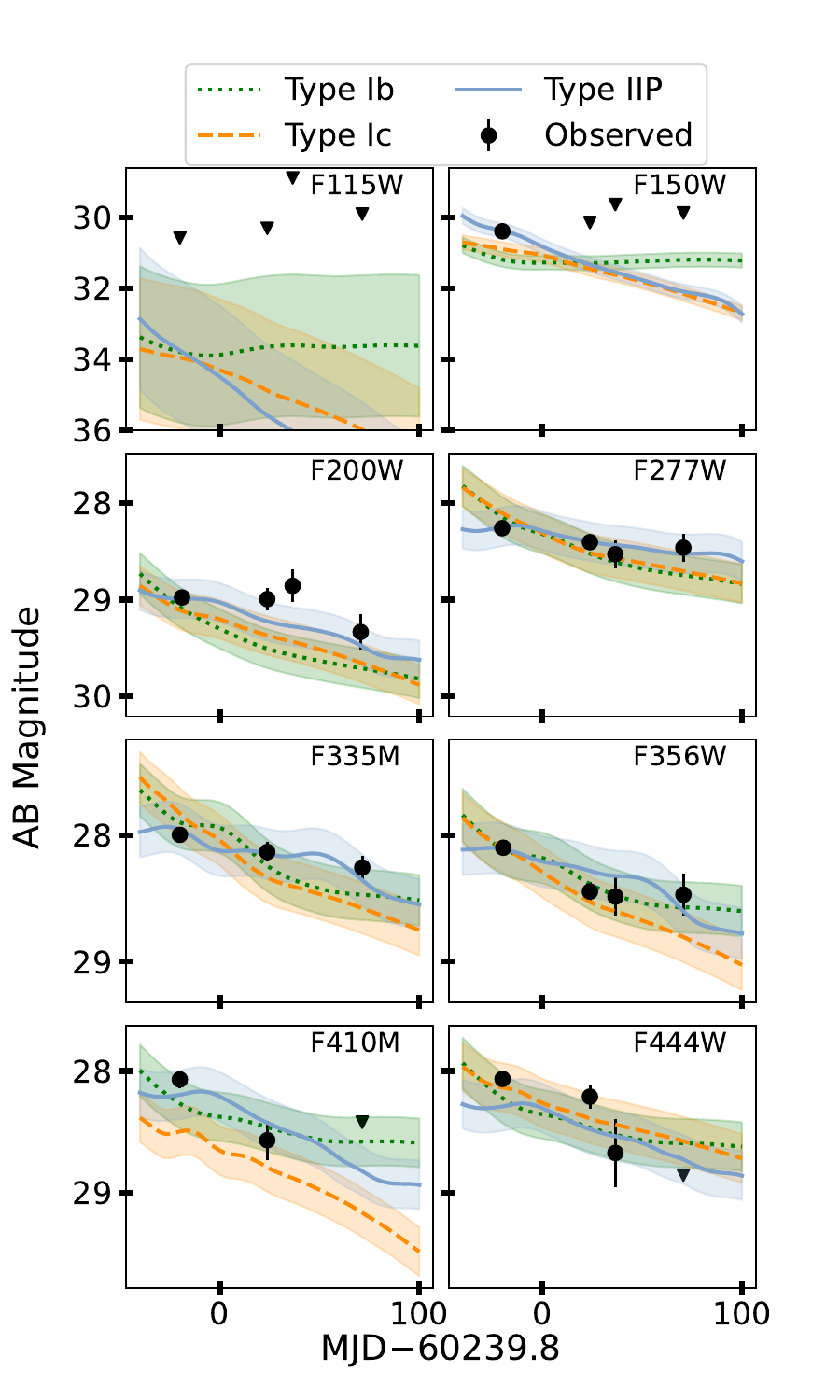}
    \caption{The photometry measured in Section \ref{sub:obs_phot} is shown as black circles with error, with ($5\sigma$) upper-limits denoted by triangles. The best-fit SN\,Ib (red dotted line), SN\,Ic (orange dashed line), and SN\,IIP (blue solid line) models are shown for comparison. Unlike SALT3-NIR these spectral templates do not have a defined model covariance, and so the uncertainties are purely statistical.}
    \label{fig:lc_fit_cc}
\end{figure}

Figures \ref{fig:lc_fit} and \ref{fig:lc_fit_cc} show the best-fit models for each SN sub-type in all filters. The resulting reduced-$\chi^2$ and measured time of peak B-band brightness for each model is shown in Table \ref{tab:phase} alongside the results from the spectroscopic analysis in Section \ref{sub:class_spec}. The SN\,Ib and SN\,Ic sub-types are heavily disfavored (best-fit $\chi^2/\nu=4.86$ and $6.00$, respectively) compared to SN\,Ia ($\chi^2/\nu=0.95$). The SN\,IIP model is a reasonable fit to the data ($\chi^2/\nu=1.34$), but the measured time of peak B-band brightness is $26\sigma$ lower than that inferred by spectral template matching (Table \ref{tab:phase}). This corresponds to a difference between light curve and spectral fits of $\sim20$ rest-frame days, while the SN\,Ia time of peak measurements from the light curve and spectrum agree within $2\sigma$ ($\sim1$ rest-frame day). Therefore the spectroscopic and photometric classification work gives consistent results only for SN\,Ia, both suggesting that we are seeing a SN\,Ia evolving from $\sim-5$ to $+20$ rest-frame days relative to peak brightness with the spectrum taken at the end of this range (Tables \ref{tab:phot} and \ref{tab:phase}). We also note that the CC\,SN $\chi^2/\nu$ values in Table \ref{tab:phase} are from the best-fit models, while the distributions for all templates for SN\,Ib, SN\,Ic, SN\,IIP are $5.82\pm0.66$, $7.43\pm0.39$, and $3.78\pm0.43$ respectively. 

\begin{figure*}
    \centering
    \includegraphics[width=\textwidth,trim={.5cm 1.5cm .5cm 3cm},clip]{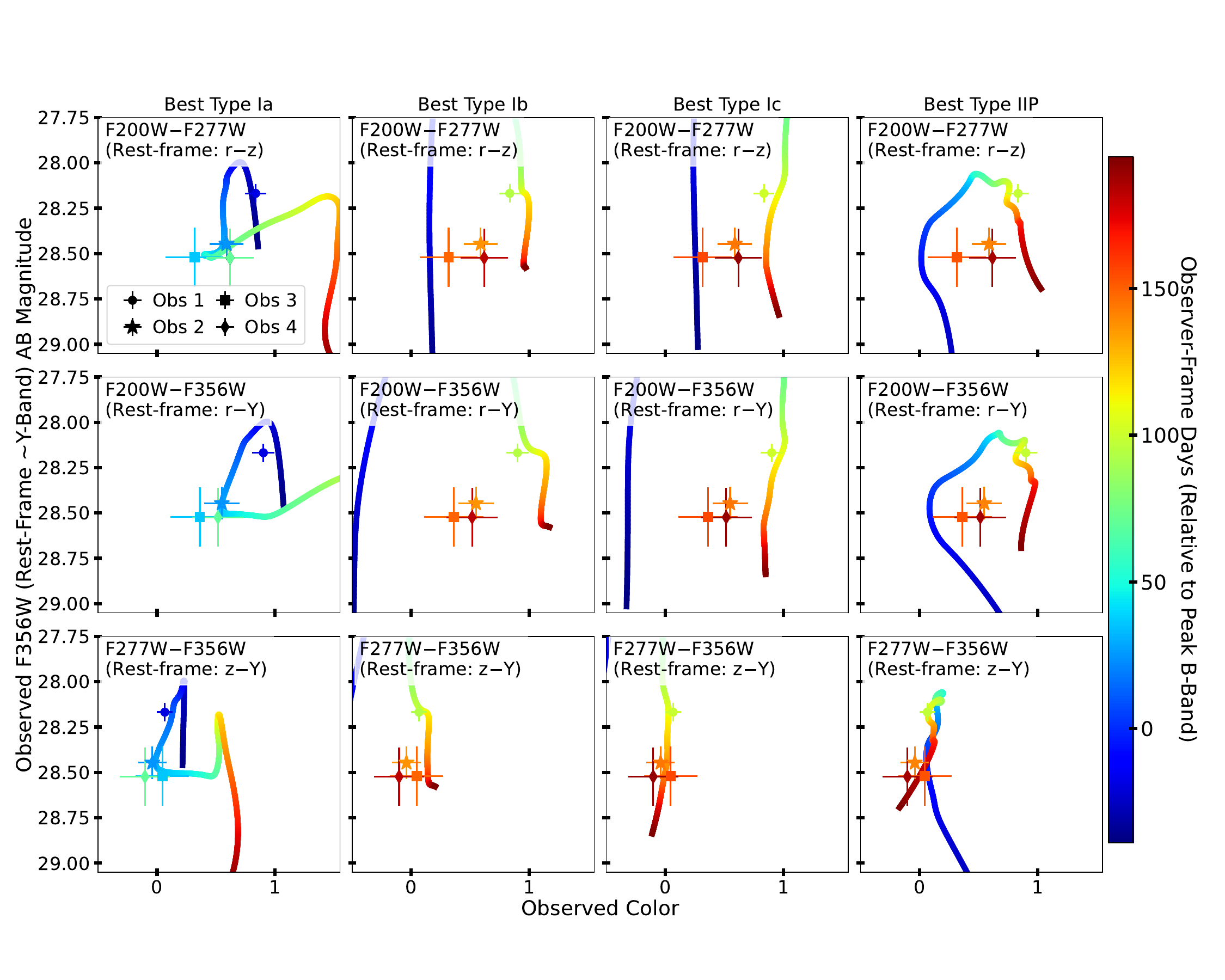}
    \caption{Three observed colors (labeled by row) vs. magnitude (F356W, rest-frame ~Y-band) shown as black points with error bars, with the symbols corresponding to the four observed epochs (legend in upper-left; order of observations is circle, star, square, diamond). The colored lines track the corresponding color-magnitude space as a function of time from best-fit models, with SN\,Ia in the left column (see Section \ref{sub:distance_lc}) and the top three CC\,SN model fits in the remaining columns. The coloring of the lines is described by the colorbar (right), with early times shown as blue and late times as red. The SN\,Ia model is the only one to accurately represent the observed color-magnitude relationships in the near-IR as a function of time.   }
    \label{fig:color_mag}
\end{figure*}

Finally, we turn to the rest-frame near-infrared (near-IR) photometry, where SNe\,Ia have a distinct second maximum that should differentiate the SN sub-classes \citep[e.g.,][]{pierel_salt3nir_2022,mandel_hierarchical_2022}. Figure \ref{fig:color_mag} shows the observed rest-frame near-IR colors vs. rest-frame near-IR magnitude for \highzIa compared to the best-fit SN\,Ia and CC\,SN models. We restrict the comparison to rest-frame rzY filters, where the light curve models are most robust. The evolution in rest-frame near-IR color-magnitude space is well-matched by the SN\,Ia template, while the CC\,SN templates fail to reproduce the observed trends as accurately. The SN\,Ic and SN\,IIP models are the next best matches in color-magnitude space, but SN\,Ic is ruled out by the overall much poorer light curve fit ($\chi^2/\nu=7.20$) and SN\,IIP by the combination of poor spectral match $(\chi^2/\nu=2.24)$ and large discrepancy between time of peak B-band brightness inferred from light curve and spectral fitting. We therefore conclude that the combination of imaging and spectroscopy for \highzIa is sufficient to classify \highzIa as a likely SN\,Ia at $z=\snz$.

\begin{table}
    \centering
    \caption{\label{tab:phase} The time of peak B-band brightness ($t_{pk}$) inferred from the light curve fitting compared to the best spectral template match for each SN type. }
    
    \begin{tabular*}{\linewidth}{@{\extracolsep{\stretch{1}}}*{5}{c}}
\toprule
SN Type&\multicolumn{2}{c}{Light Curve}&\multicolumn{2}{c}{Spectroscopic}\\
&$t_{pk}$&$\chi^2/\nu$&$t_{pk}$&$\chi^2/\nu$\\
\hline
Ia&$60240\pm2$&$0.95$&$60236$&$1.72$\\
Ib&$60127\pm4$&$4.86$&$60138$&$2.47$\\
Ic&$60120\pm4$&$6.00$&$60232$&$1.92$\\
IIP&$60123\pm3$&$1.34$&$60201$&$2.24$\\

\end{tabular*}
\begin{flushleft}
\tablecomments{Columns are: SN type model/spectral template used, the time in Modified Julian Date (MJD) of peak B-band brightness measured from light curve fitting (to the data $<4\mu$m where models are better constrained), the light curve fitting $\chi^2$ per degree of freedom (DOF; $\nu$) \textit{without} model uncertainties, as they do not exist for CC\,SN models, the time of peak B-band brightness given the best-fit spectroscopic template match, and the $\chi^2$ per DOF of the best-fit spectroscopic template match.}

\end{flushleft}
\end{table}

\subsection{Comparison to low-$z$ Ca-rich SNe\,Ia}
\label{sub:class_ca-rich}
Using the full wavelength range of the spectrum, the best-match spectral template for \highzIa from \texttt{NGSF} is a normal SN\,Ia despite the presence Ca-rich templates in the database. We turn to the population of Ca-rich SNe\,Ia to explain the high observed Ca\,II velocity and red color, but note that \highzIa appears to best-match a normal SN\,Ia apart from these characteristics. We find that the Ca\,II velocity ($\sim18,000$km s$^{-1}$ and red intrinsic color ($E(B-V)\sim1$ near peak brightness) of SN \,2016hnk, both measured by \citet{galbany_evidence_2019} and \citet{jacobson-galan_ca_2020}, is the best match to \highzIa ($\sim19,000$km s$^{-1}$ and ($E(B-V)\sim0.9$). We show a comparison of SN\,2016hnk at $\sim10$ days after peak brightness to the \highzIa spectrum in Figure \ref{fig:sn_spec}, and focus on the Ca\,II feature in Figure \ref{fig:sn_spec_ca}. The match is quite good despite \texttt{NGSF} preferring a normal SN\,Ia, suggesting that \highzIa may share some properties with Ca-rich transients \citep[Both observed and theoretical:][]{woosley_models_1986,bildsten_faint_2007,perets_faint_2010,shen_thermonuclear_2010,waldman_helium_2011,kasliwal_calcium-rich_2012,foley_kinematics_2015,de_zwicky_2020,zenati_origins_2023}. We note that while the color at peak B-band brightness and Ca\,II velocity seem to match well between these two objects, the absolute B-band magnitude of \highzIa is $\sim1$mag brighter than SN\,2016hnk before standardization. This puts \highzIa more in the luminosity range of $91$bg-like SNe\,Ia \citep{filippenko_subluminous_1992,taubenberger_underluminous_2008,sullivan_subluminous_2011,taubenberger_extremes_2017}, but unlike the fast-declining $91$bg-like SNe\,Ia our fits to \highzIa are consistent with a normal decline rate (see Section \ref{sub:distance_lc}). More SNe\,Ia in this new redshift range are needed to determine if \highzIa is peculiar, or if very high-$z$ SNe\,Ia typically share properties with both normal and less common SN\,Ia sub-types.

\section{Luminosity Distance Measurement}
\label{sec:distance}

\subsection{Light Curve Fitting}
\label{sub:distance_lc}

\begin{table}
    \centering
    \caption{\label{tab:fit} The SALT3-NIR light curve model parameters used in this analysis.}
    
    \begin{tabular*}{\linewidth}{@{\extracolsep{\stretch{1}}}*{3}{c}}
\toprule
Parameter&Bounds&Best-Fit\\
\hline
$z$&Fixed&$z=\snz$\\
$t_{pk}$&[60200,60320]&$60239.98^{+1.50}_{-1.70}$\\
$x_0$&[0,1]&$(7.12_{-0.55}^{+0.66})\times10^{-9}$\\
$x_1$&[-3,3]&$-0.11^{+1.03}_{-1.06}$\\
$c$&[-1.5,1.5]&$0.92^{+0.04}_{-0.05}$
\end{tabular*}
\end{table}
We begin by fitting the observed photometry (including upper-limits; Table \ref{tab:phot}) with the SALT3-NIR SN\,Ia light curve evolution model \citep{pierel_salt3nir_2022}, which has rest-frame wavelength coverage of $\sim2,500$-$20,000\AA$. In addition to the basic light curve parameters of redshift, amplitude ($x_0$), and time of peak brightness ($t_{pk}$), SALT3-NIR parameterizes SN\,Ia light curves with the ``shape'' or ``stretch'' ($x_1$) and color ($c$) parameters. These are used in Section \ref{sub:distance_hd} to make the traditional corrections to the observed peak apparent magnitude needed to obtain a standardized luminosity distance \citep[e.g.,][]{tripp_two-parameter_1998}. We include the same $0.01$mag Galactic extinction correction as in Section \ref{sub:class_phot}, and allow for high values for the SALT3-NIR color parameter (up to $c=1.5$). For this stage, it is critical that the measured $x_1$ and $c$ values are accurate, but including the poorly constrained SALT3-NIR rest-frame J-band leads to biased parameters when included in the fit (Red model, Figure \ref{fig:lc_fit}). We therefore remove the F410M and F444W filters from the fit for an accurate standardization, resulting in the blue model in Figure \ref{fig:lc_fit}. The bounds and retrieved SALT3-NIR parameters from the fit are shown in Table \ref{tab:fit}. 

The best-fit model is shown with the observed photometry in Figure \ref{fig:lc_fit}. SALT3-NIR is an excellent fit in all filters below $4\mu$m and matches the $F410M/F444W$ filters within $1\sigma$, but is systematically fainter than these reddest filters whether they are included in the fit or not. This could be due to the large uncertainties in the rest-frame J-band model \citep{pierel_salt3nir_2022}, an issue with the PSF model used at these reddest wavelengths, and/or a zero-point offset in the $4\mu$m data. Regardless, when the $4\mu$m data are included, the fitter attempts to vary the model parameters to an extreme degree in order to improve the fit, biasing the results. Since the resulting fit is not a large improvement in the $>4\mu$m filters and degrades the fit at $<4\mu$m, we proceed with the fit to data with wavelengths $<4\mu$m (rest-frame $\leq1\mu$m).

\subsection{Simulations for Bias Correction}
\label{sub:distance_bias}
We simulate the discovery JADES epoch using the Supernova Analysis (SNANA) code \citep{kessler_snana:_2009,kessler_first_2019} to make an approximate correction for bias from selection effects, Malmquist bias, and light curve fitting bias from our luminosity distance measurement for \highzIa. SNANA simulates SN light curves for an arbitrary set of survey properties while accounting for variations in noise, PSF, and cadence. Due to its speed, accuracy, and flexibility, SNANA has become the standard tool for simulating SN surveys in recent years \citep[e.g.,][]{betoule_improved_2014,scolnic_complete_2018,jones_foundation_2019,kessler_first_2019,rose_reference_2021,brout_pantheon_2022}. Following Figure 1 in \citet{kessler_first_2019}, a brief overview of the SNANA simulation scheme that we apply to this analysis is as follows:
\begin{enumerate}
    \item \textbf{Source Model}
    \begin{enumerate}
        \item Generate source SED at each simulated epoch using SALT3-NIR.  We use SN\,Ia parameter distributions from \citet{popovic_pantheon_2023}, but extend the color range out to $c = 1.1$ to match the particularly red color of this SN.
        \item Apply cosmological dimming, Galactic extinction, weak lensing,  and redshift to the SED.  We simulate every SN at the \highzIa redshift of $\snz$ as there is negligible uncertainty in the redshift.
        \item Integrate the redshifted SED over each filter transmission function to create the noise-free photometric light curve.
    \end{enumerate}
    \item \textbf{Noise Model}
    \begin{enumerate}
        \item Use image zero-point to convert each true light curve in magnitude 
    to true flux in 
    photoelectrons.
        \item Compute flux uncertainty from zero-point, PSF and sky noise, which are determined on a per-epoch basis from the real JADES observations. These uncertainties are used to apply Gaussian-random fluctuations to true fluxes.
    \end{enumerate}
    \item \textbf{Trigger Model}
    \begin{enumerate}
        \item Check for detection (S/N $>3\sigma$ in 2 or more bands).
        \item Write selected events to data files.
    \end{enumerate}
\end{enumerate}

We simulated a sample of 20,000 SNe\,Ia and fit the full sample with SALT3-NIR.  After fitting, we then select the SN sample that has best-fit parameters that closely match the real best-fit parameters $x_1$, and $c$.  We assume the true $\alpha$ and $\beta$ are equal to those measured by the Pantheon$+$ team \citep{brout_pantheon_2022}, discussed in Section \ref{sub:distance_hd}.  The $\alpha$ and $\beta$ values were estimated from a large SN sample and therefore their uncertainties should have a negligible impact on the uncertainties for a single SN (though $z$-dependent evolution in these parameters is a potential concern that we do not address in this work).  For that simulated set, the bias correction is the average difference between the Tripp-derived distance modulus when the {\it fitted} parameters are used versus the Tripp-derived distance modulus when the {\it simulated} parameters are used.

This method is an approximation of the BEAMS with Bias Corrections (BBC) method \citep{kunz_bayesian_2007,kunz_beams_2013,kessler_correcting_2017}, which estimates a correction term $\Delta\mu_{\rm{Bias}}$ from the difference between simulated versus recovered parameters, based on a large simulated sample of SNe\,Ia in a 5D space of \{$z,x_1,c,\alpha,\beta$\} (see Section \ref{sub:distance_hd}). 

The methods follow those of previous cosmological analyses \citep[e.g.,][]{scolnic_complete_2018} with the modest simplifications described above due to having just a single SN instead of hundreds to thousands.  

From this approach, we find that the bias correction is fairly negligible at $\sim2$\%, though we still include it in the final luminosity distance measurement (Section \ref{sub:distance_hd}).  This implies that nearly all normal SNe\,Ia within the observed range of $x_1$,$c$ values would be detected by our survey at this redshift, due to the extreme depth excellent wavelength coverage provided by the JADES program (Table \ref{tab:phot}).

\begin{figure*}[ht!]
    \centering
    \includegraphics[width=\textwidth,trim={0cm 0cm 2.5cm 1.5cm},clip]{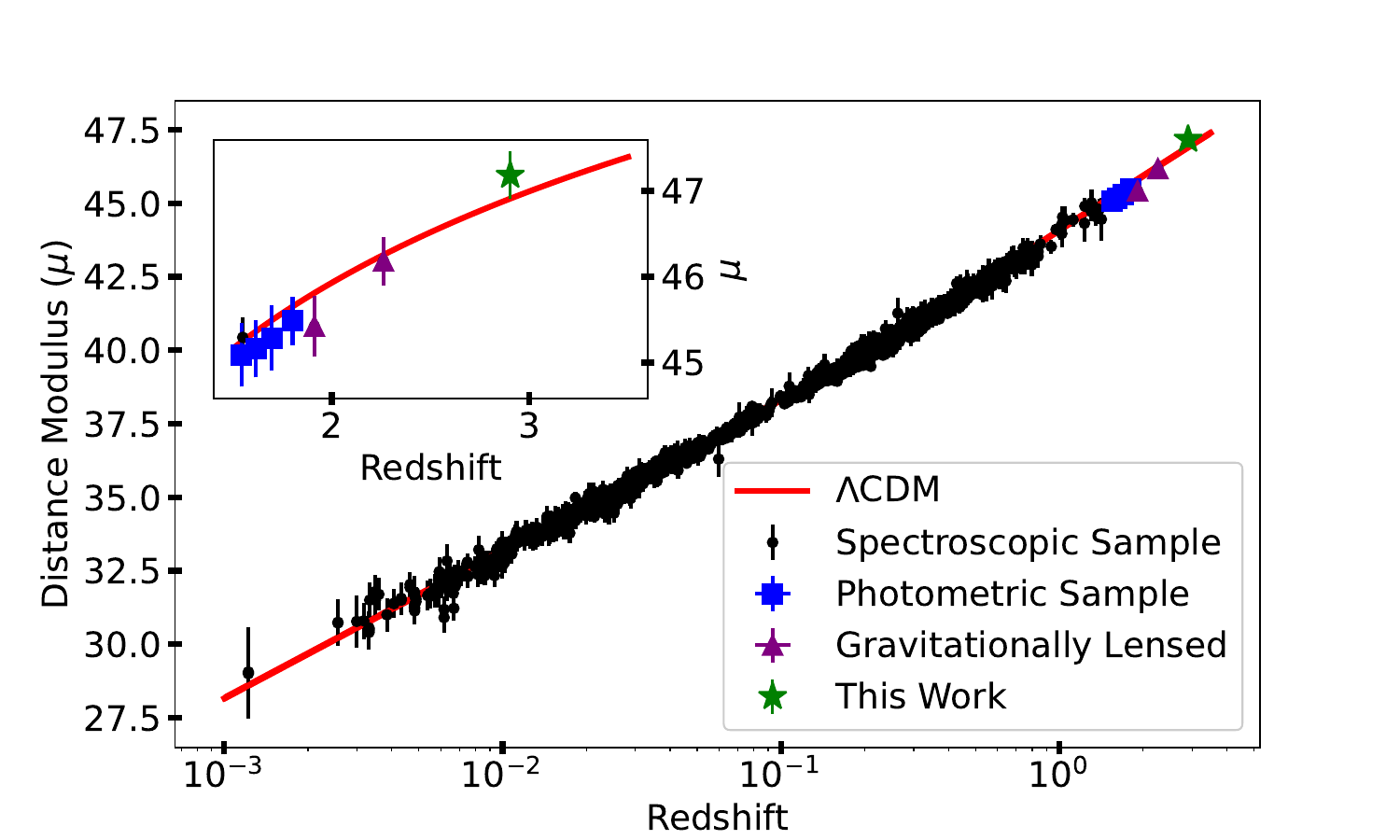}
    \caption{Luminosity distance measurements from the full sample of SNe\,Ia from \citet{brout_pantheon_2022} extending to $z=2.22$. Black points (with errors) are SNe\,Ia with spectroscopic classifications, while blue squares (with error) are SNe\,Ia with photometric classifications. The two gravitationally lensed SNe\,Ia with distance measurements are shown as purple triangles. \highzIa is shown as a green star, and $\Lambda$CDM is shown as a solid red line for reference. The width of the red line encompasses the width of the current $H_0$ tension, with the center of the line used for reference. }
    \label{fig:lcdm}
\end{figure*}

\subsection{Extending the Hubble Diagram to $z=3$}
\label{sub:distance_hd}
We transform fitted SALT3-NIR light curve parameters from Section \ref{sub:distance_lc} into a distance by way of a modified Tripp formula \citep{tripp_two-parameter_1998}:
\begin{equation}
\label{eq:tripp}
\mu=m_B-M+\alpha x_1-\beta c+\delta_{host}+\Delta\mu_{\rm{Bias}},
\end{equation}
where $\mu$ is the distance modulus, $m_B$ is the peak apparent magnitude in the rest-frame B-band, $\alpha$ ($\beta$) is the coefficient of relation between SN\,Ia luminosity and stretch (color), and $M$ is the peak absolute magnitude of an $x_1=c=0$ SN\,Ia assuming some nominal value of $H_0$ (here $H_0=70$ km s$^{-1}$ Mpc$^{-1}$ and $M=-19.36$). The $\delta_{host}$ parameter is the host-galaxy mass step, or the small residual correlation between SN\,Ia distance measurements and their host-galaxy masses \citep{kelly_hubble_2010,lampeitl_effect_2010,sullivan_dependence_2010}; because the nature and evolution of the host-galaxy mass step is unknown, especially at such high redshift \citep[e.g.,][]{childress_ages_2014}, we simply apply half of the host mass step (for a low-mass galaxy, see Section \ref{sec:conclusion}) from \citet{brout_pantheon_2022} (who found $\sim0.054$mag using the same scatter model implemented in Section \ref{sub:distance_bias}) and add a systematic error of half the host mass step in quadrature. Finally, the $\Delta\mu_{\rm{Bias}}$ term is a selection bias correction determined by BBC, described in Section \ref{sub:distance_bias}, which we constrain to be $-0.018$mag for this analysis. Without a large sample of high-$z$ SNe to measure the nuisance parameters ($\alpha$, $\beta$), we fix $\alpha=0.148$ and $\beta=3.09$ \citep[these parameters do not seem to change with redshift, but more high-$z$ SNe\,Ia are needed to confirm the result;][]{scolnic_complete_2018}, which are the best constraints from $z\lesssim2$ SNe\,Ia by \citet{brout_pantheon_2022}. The $m_B$ parameter found by the SALT3-NIR model to be $30.73$, and the shape/color parameters are shown in Table \ref{tab:fit}. 

The final luminosity distance measurement is $\mufit_{-\mufiterrl}^{+\mufiterrh}$mag, while the $\Lambda$CDM prediction at $z=\snz$ (with $H_0=70$ km s$^{-1}$ Mpc$^{-1}$) is $\mu=46.91$mag, a $\lesssim1\sigma$ difference (Figure \ref{fig:lcdm}). The uncertainty on $\mu$ includes the fitted model uncertainties, errors from redshift and peculiar velocity (which are negligible here), the intrinsic scatter of SNe\,Ia \citep[$0.1$mag;][]{scolnic_complete_2018}, and an additional $0.005z$\,mag uncertainty from weak gravitational lensing \citep{jonsson_constraining_2010}. We note that \highzIa would not pass fiducial cosmological cuts \citep[$|c|<0.3$;][]{scolnic_complete_2018} because of its red color ($c\sim0.9$), but applying the traditional standardization nevertheless results in this agreement with $\Lambda$CDM. More high-$z$ SNe\,Ia are required to determine if there is true drift in the normal SN\,Ia population parameters, meaning high-$z$ SNe\,Ia could be intrinsically redder than low-$z$ SNe\,Ia while still adhering to a normal Tripp equation for standardization, or if this object is peculiar for its redshift.

\section{Discussion}
\label{sec:conclusion}
We have presented \textit{JWST} observations of a SN (\highzIa) with a spectroscopic redshift of $z=\snz\pm0.007$, which we classify using both the spectrum and light curve information as the most distant SN\,Ia yet discovered. We note that \highzIa could plausibly still be a CC\,SN that appears different from our finite low-$z$ library of spectra and light curve models, but a larger sample of high-$z$ CC\,Se observed light curve with a model for SN\,Ia evolution that includes a traditional shape and color parameterization and use our current best understanding of SN\,Ia standardization at lower redshift ($z\lesssim2$) to measure the luminosity distance to \highzIa. Although \highzIa would not pass fiducial low-$z$ cosmology cuts because of its red color we find a value of $\mu=\mufit_{-\mufiterrl}^{+\mufiterrh}$mag including a correction for potential observational biases, which is in excellent agreement ($\lesssim1\sigma$) with $\Lambda$CDM. Although a single object is not enough to directly constrain cosmological parameters at high-$z$, any significant deviation of SN\,Ia luminosity distances from $\Lambda$CDM at $z>2$ would be a strong indicator of SN\,Ia luminosity evolution with redshift. The agreement of \highzIa with $\Lambda$CDM, the most distant such test, gives no indication of significant SN\,Ia luminosity evolution with redshift. 

Despite agreement between the \highzIa luminosity distance measurement and standard cosmology, there are two observed peculiarities with \highzIa. The first is its very red observed color ($c\sim0.9$), and the second is its high Ca\,II velocity ($\sim19000$km$^{-1}$). The red color could be attributed to significant dust attenuation from the host galaxy JADES-GS$+53.13485$$-$$27.82088$\footnote{JADES Host ID 96906 from \url{https://archive.stsci.edu/hlsp/jades}}, but fitting photometry of the host galaxy from $2022$ \citep[well before the SN explosion;][]{eisenstein_jades_2023} with the Bayesian Analysis of Galaxies for Physical Inference and Parameter EStimation \citep[\texttt{Bagpipes};][]{carnall_inferring_2018} infers a fairly low-mass ($\sim10^{8}M_\odot$), low-metallicity ($\sim0.3Z_\odot$), low-extinction ($A_V<0.1$) host galaxy, suggesting that \highzIa could be intrinsically red. Low-$z$ SNe with high Ca\,II velocities tend to be redder than the general population of SNe\,Ia \citep{siebert_investigating_2019}, but \highzIa is still fairly extreme in both parameters. The low-$z$ SN 2016hnk is a good match to both the \highzIa color and Ca\,II velocity, but has a $\sim1$mag fainter absolute magnitude before standardization. \highzIa has an absolute magnitude closer to $91$bg-like SNe\,Ia, but with a normal light curve decline rate. We require a larger population of high-$z$ SNe\,Ia to determine if \highzIa is truly an outlier that should be cut from future cosmological constraints (i.e., most normal high-$z$ SNe\,Ia fall within traditional low-$z$ cosmology cuts) or if the distribution of SN\,Ia properties varies significantly with redshift due to changes in progenitors or their environment. 

\highzIa is the first SN\,Ia candidate with a combined spectroscopic and photometric dataset in the dark matter dominated universe at $z>2$, making this the first robust test for SN\,Ia standardized luminosity evolution in the manner suggested by \citet{riess_first_2006}. \textit{JWST} is the only resource capable of expanding this sample further, and is expected to do so with $\gtrsim10$ additional such objects anticipated over the next two years \citep{pierel_pass_2024}. While \highzIa gives no indication that standardized SN\,Ia luminosities evolve significantly with redshift, the full sample will be required to confirm this result and put constraints on any possible evolution at lower redshift for future cosmological measurements.

\begin{center}
    \textbf{Acknowledgements}
\end{center}

We would like to thank Erin Hayes, Saurabh Jha and Rick Kessler for useful discussion. This paper is based in part on observations with the NASA/ESA Hubble Space Telescope and James Webb Space Telescope obtained from the Mikulski Archive for Space Telescopes at STScI. We thank the DDT and JWST/HST scheduling teams at STScI for extraordinary effort in getting the DDT observations used here scheduled quickly. This work is based on observations made with the NASA/ESA/CSA James Webb Space Telescope. The data were obtained from the Mikulski Archive for Space Telescopes at the Space Telescope Science Institute, which is operated by the Association of Universities for Research in Astronomy, Inc., under NASA contract NAS 5-03127 for JWST. These observations are associated with program \#1180 and 6541.
This research is based (in part) on observations made with the NASA/ESA Hubble Space Telescope obtained from the Space Telescope Science Institute, which is operated by the Association of Universities for Research in Astronomy, Inc., under NASA contract NAS 5–26555. Part of the JWST data used in this paper can be found in MAST: \dataset[10.17909/8tdj-8n28]{https://dx.doi.org/10.17909/8tdj-8n28} (JADES DR1). 
Additionally, this work made use of the {\it lux} supercomputer at UC Santa Cruz which is funded by NSF MRI grant AST 1828315, as well as the High Performance Computing (HPC) resources at the University of Arizona which is funded by the Office of Research Discovery and Innovation (ORDI), Chief Information Officer (CIO), and University Information Technology Services (UITS).
AJB acknowledges funding from the “FirstGalaxies” Advanced Grant from the European Research Council (ERC) under the European Union’s Horizon 2020 research and innovation program (Grant agreement No. 789056). 
PAC, EE, DJE, BDJ, are supported by JWST/NIRCam contract to the University of Arizona, NAS5-02015.
DJE is also supported as a Simons Investigator.
RM acknowledges support by the Science and Technology Facilities Council (STFC), by the ERC through Advanced Grant 695671 “QUENCH”, and by the UKRI Frontier Research grant RISEandFALL. RM also acknowledges funding from a research professorship from the Royal Society.
BER acknowledges support from the NIRCam Science Team contract to the University of Arizona, NAS5-02015, and JWST Program 3215. JDRP is supported by NASA through a Einstein
Fellowship grant No. HF2-51541.001 awarded by the Space
Telescope Science Institute (STScI), which is operated by the
Association of Universities for Research in Astronomy, Inc.,
for NASA, under contract NAS5-26555.

\clearpage

\bibliographystyle{aasjournal}


\end{document}